\documentclass[prd,aps,twocolumn,showpacs,preprintnumbers,amsmath,amssymb]{revtex4}
\usepackage[dvips]{graphicx}
\usepackage{epsf}
\usepackage{amsmath}
\usepackage{amssymb}

\usepackage{graphicx}
\usepackage{dcolumn}
\usepackage{bm}
\def\be{\begin{equation}}
\def\ee{\end{equation}}
\def\bea{\begin{eqnarray}}
\def\eea{\end{eqnarray}}

\newcommand{\comment}[1]{}

\newcommand{\tphi}{\tilde{\phi}}

\begin{document}


\date{\today}

\title{Matter Bounce in Ho\v{r}ava-Lifshitz Cosmology}

\author{Robert Brandenberger$^{1,2}$}

\affiliation{1) Department of Physics, McGill University,
Montr\'eal, QC, H3A 2T8, Canada}

\affiliation{2) Theory Division, CERN, CH-1211 Geneva, Switzerland}


\begin{abstract}

Ho\v{r}ava-Lifshitz gravity, a recent proposal for a UV-complete renormalizable
gravity theory, may lead to a bouncing cosmology. In this note we argue that
Ho\v{r}ava-Lifshitz cosmology may yield  a concrete realization of the matter bounce
scenario, and thus give rise to an alternative to inflation for producing a scale-invariant
spectrum of cosmological perturbations. In this scenario, quantum vacuum
fluctuations exit the Hubble radius in the pre-bounce phase and the
spectrum is transformed into a scale-invariant one on super-Hubble
scales before the bounce because the long wavelength modes undergo
squeezing of their wave-functions for a longer period of time than shorter
wavelength modes. The scale-invariance of the spectrum of curvature
fluctuations is preserved during and after the bounce. A distinctive prediction
of this scenario is the amplitude and shape of the bispectrum. 

\end{abstract}

\maketitle

\newcommand{\eq}[2]{\begin{equation}\label{#1}{#2}\end{equation}}

\section{Introduction}

Recently, Ho\v{r}ava (based on the pioneering work of \cite{Horava0}
has proposed \cite{Horava1,Horava2} a model for quantum
gravity which is power-counting renormalizable and hence 
potentially ultra-violet (UV) complete.
This model does not have the complete diffeomorphism invariance
of General Relativity, but the action has a fixed point in the infrared (IR)
which is that of General Relativity with a negative cosmological constant.
In the UV, however, the theory flows to a different fixed point, a fixed
point at which space and time scale differently and
which has much better UV behavior of perturbation theory. Since
Ho\v{r}ava's theory is modelled after a scalar field model studied by
Lifshitz \cite{Lifshitz} in which the full Lorentz symmetry also only emerges
at an IR fixed point, the theory is now called Ho\v{r}ava-Lifshitz gravity.

Specific solutions of the simplest version of Ho\v{r}ava-Lifshitz gravity have recently been
analyzed. In \cite{Soda}, homogeneous vacuum solutions with gravitational
waves were studied. in \cite{Calcagni,Kiritsis}, cosmological solutions with
matter were explored, and in \cite{Pope}, black hole solutions were analyzed,
As pointed out in \cite{Calcagni,Kiritsis}, the analogs of the Friedmann
equations in Ho\v{r}ava-Lifshitz gravity include a term which scales as dark
radiation and contributes a negative term to the energy density. Thus,
it is possible in principle to obtain a nonsingular cosmological evolution
with the Big Bang of Standard and Inflationary Cosmology replaced by
a bounce.

In \cite{Horava1,Kiritsis}, it was argued that the different ultraviolet behavior of the
theory might provide an alternative to cosmological inflation for solving
the problems of Standard Cosmology such as the horizon and flatness
problems. Specifically, the divergence of the speed of light in the far
ultraviolet leads to the possibility of solving the horizon problem as
proposed a while back in \cite{Moffat,Joao}. In \cite{Calcagni} it
was emphasized that if the wavelength of fluctuations penetrates
the UV region, then the usual arguments for the origin of a scale-invariant
spectrum of cosmological perturbations from an inflationary phase
might break down, as suggested more generally in the context of
the ``trans-Planckian problem" for inflationary fluctuations
\cite{RHBrev,Jerome}.

However, if Ho\v{r}ava-Lifshitz cosmology leads to a cosmological
bounce, then it is not necessary to invoke a period of inflationary
expansion to produce the observed spectrum of cosmological
perturbations. The purpose of this Note is to point out that
Ho\v{r}ava-Lifshitz cosmology may provide a UV-complete
realization of the ``matter bounce" scenario (see \cite{RHBrev2} for
an introduction to this scenario), an alternative to cosmological
inflation for explaining the origin of the observed structure in
the Universe.

As realized in \cite{Wands,FB2,Wands2}, perturbations which 
start out as quantum vacuum fluctuations and exit the
Hubble radius during a matter-dominated phase of contraction
acquire a scale-invariant spectrum. Given a non-singular
bouncing background cosmology, the fluctuations can be
followed unambiguously through the bounce. If the energy
density at which the bounce occurs is smaller than the Planck
scale, then the wavelength of the fluctuations which are being
probed in today's observations are in the far IR (they are
a fraction of a ${\rm mm}$). Hence, the equations which
describe these fluctuations are those of the IR fixed point
of the theory, which is the Einstein action in the case of
Ho\v{r}ava-Lifshitz cosmology. It has been shown 
\cite{Omidbounce,Tirthobounce,quintombounce,LWbounce}
that, provided the duration of the bounce phase is short
compared to the wavelength of the fluctuations being
considered, then the spectrum of curvature fluctuations
is not changed during the bounce. Thus, a scale-invariant
spectrum of curvature perturbations will persist in the
post-bounce expanding phase. Specific predictions
of the matter bounce scenario include a specific form
on the non-Gaussianities as measured by the amplitude
and shape of the bispectrum \cite{matterbounceng}. 

In this Note we give an overview of the Ho\v{r}ava-Lifshitz
matter bounce cosmology, leaving details for future
investigations. We begin with a space-time sketch
depicting the relevant phases (Figure 1). The vertical
axis is time, with $t = 0$ denoting the bounce time.
In some early phase of contraction, the equation of
state of matter is assumed to be dominated by
non-relativistic pressure-less matter, in the same sense that our
current expanding universe is. For times between
$- t_m$ and the bounce, the equation of state 
can be different from that of pressure-less matter. The horizontal
axis denotes co-moving spatial coordinates. Vertical lines
correspond to fixed co-moving wavelengths, the dashed
line is the co-moving Hubble radius $H^{-1}$. Fluctuations
which cross the Hubble radius during the matter phase
of contraction acquire a scale-invariant spectrum, those
which cross later have a non-trivial spectral slope whose
magnitude depends on the specific equation of state
(see e.g. \cite{Hongli}).

The outline of this Note is as follows. We first review
the action of Ho\v{r}ava-Lifshitz gravity.  Next, we review
the equations for cosmological solutions and study
the possibility of obtaining a bouncing cosmology
with a matter-dominated phase of contraction.
In Section 4 we review the evolution of fluctuations
in the contracting phase of the matter bounce scenario,
and in Section 5 we study how fluctuations pass through
the bounce in Ho\v{r}ava-Lifshitz cosmology. We
conclude with a discussion of some of the many open
issues.

\begin{figure}[htbp]
\includegraphics[scale=0.5]{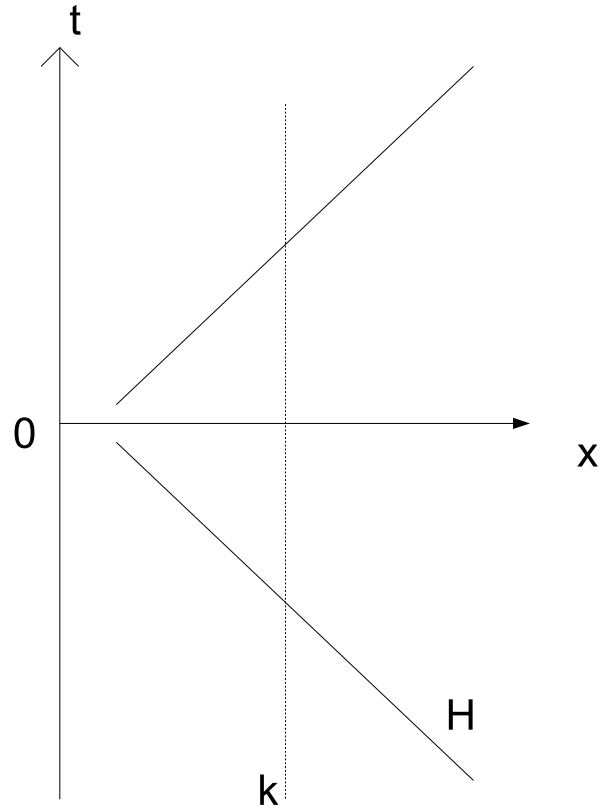}
\caption{A space-time sketch of the matter bounce scenario. The vertical
axis is time, with $t = 0$ being the bounce time. The horizontal axis
denotes co-moving distance. The curve with the label $H$ is the Hubble radius
$H^{-1}$ - in co-moving coordinates, the vertical line labeled by $k$
denotes the co-moving wavelength of a fluctuation mode. This mode
crosses the Hubble radius in the contracting phase before the time $-t_c$,
when the period of matter-domination ends. }
\end{figure}

\section{Review of Ho\v{r}ava-Lifshitz Gravity }

We begin with a brief review of Ho\v{r}ava-Lifshitz
gravity \footnote{This section is based
completely on the analysis of \cite{Kiritsis}.}
The dynamical variables are the
lapse and shift functions $N$ and $N_i$, respectively, and
the spatial metric $g_{ij}$ (roman letters indicate spatial
indices) In terms of these fields, the full
metric is
\be
ds^2 \, = \, - N^2 dt^2 + g_{ij} \bigl(dx^i + N^i dt \bigr)
\bigl( dx^j + N^j dt \bigr) \, ,
\ee
where the indices of $N$ are raised and lowered using the
spatial metric $g_{ij}$. 

The scaling symmetry of the coordinates in the simplest
version of Ho\v{r}ava-Liftshitz gravity is (we are following
the notation of  \cite{Kiritsis})
\be
t \, \rightarrow \, l^3 t \,\,\,\, {\rm and} \,\,\,\, x^i \, \rightarrow \, l x^i \, .
\ee

The full action of this version of Ho\v{r}ava-Lifshitz gravity is
\bea
S \, &=& \, \int dt d^3x \sqrt{g} N \bigl[ \frac{2}{\kappa^2} (K_{ij}K^{ij} - \lambda K^2) \nonumber \\
& & \, - \frac{\kappa^2}{2 w^4} C_{ij}C^{ij} 
+ \frac{\kappa^2 \mu}{2 w^2} \frac{\epsilon^{ijk}}{\sqrt{g}} R_{il} \nabla_j R^l_k \\
& & \, - \frac{\kappa^2 \mu^2}{8} R_{ij} R^{ij}
+ \frac{\kappa^2 \mu^2}{8(1 - 3 \lambda)} 
\bigl( \frac{1 - 4 \lambda}{4} R^2 + \Lambda R - 3 \Lambda^2 \bigr) \bigr]  \, , \nonumber
\eea
where 
\be
K_{ij} \, = \, \frac{1}{2N} \bigl( {\dot{g_{ij}}} - \nabla_i N_j - \nabla_j N_i \bigr) \, ,
\ee
and $C_{ij}$ is the Cotton tensor
\be
C^{ij} \, = \, \frac{\epsilon^{ijk}}{\sqrt{g}} \nabla_k \bigl( R^j_i - \frac{1}{4} R \delta^j_i \bigr) \, .
\ee
The tensor $\epsilon^{ijk}$ is the totally antisymmetric unit tensor, $\lambda$ is a dimensionless
constant and $\Lambda$ is related to the cosmological constant in the IR limit. The
variables $\kappa$, $w$ and $\mu$ are constants with mass dimensions $-1$, $0$ and $1$,
respectively. 

In the IR limit, the action reduces to 
\be
S_E \, = \, \int dt d^3x \sqrt{g} N \bigl[ \alpha (K_{ij} K^{ij} - \lambda K^2 ) 
+ \xi R + \sigma \bigr] \, ,
\ee
with
\bea
\alpha \, &=& \, \frac{2}{\kappa^2} \, \\ 
\xi \, &=& \, \frac{\kappa^2 \mu^2}{8 ( 1 - 3 \lambda)} \Lambda \,\,\, \rm{and} \\
\sigma \, &=& \, - 3 \frac{\kappa^2 \mu^2}{8 ( 1 - 3 \lambda)} \Lambda^2 \, .
\eea
In order to obtain the Einstein action, we require $\lambda = 1$. In this case,
the variables of the Ho\v{r}ava-Lifshitz action go over into the following
expressions for the speed of light $c$, Newton's gravitational constant $G$
and the effective cosmological constant $\Lambda_E$:
\bea
c \, &=& \, \sqrt{\frac{\xi}{\alpha}} \, , \\
16 \pi G \, &=& \, \sqrt{\frac{\xi}{\alpha^3}} \, , \\
\Lambda_E \, &=& \, - \frac{\sigma}{2 \alpha} \, ,
\eea

In the following we will consider scalar field matter in the contracting phase. The
action for matter is
\be
S_M \, = \, \int dt d^3x \sqrt{g} N {\cal{L}}_m \, ,
\ee
where the matter Lagrangian ${\cal{L}}_m$ depends on the scalar matter
field $\varphi$ and the metric. In the IR limit, this action reduces to the usual scalar
field matter action in curved space-time. The form of the scalar field
Lagrangian valid also in the UV is given in \cite{Kiritsis} but will not be
used in this Note.

\section{Matter Bounce in Ho\v{r}ava-Lifshitz Cosmology}

To obtain the equations for Ho\v{r}ava-Lifshitz cosmology we assume
that the metric is homogeneous and isotropic, i.e.
\be
N \, = \, N(t), \,\,\,\, N_i = 0 \,\,\,\, {\rm and} \,\,\,\, g_{ij} = a^2(t) \gamma_{ij} \, ,
\ee
where $\gamma_{ij}$ is a maximally symmetric constant curvature metric.
We will denote the spatial curvature parameter by ${\bar k}$. 

The equations of motion are obtained by varying the action with
respect to $N$, $a$, and $\varphi$, and setting $N = 1$ at the end of the
calculation. The resulting equations are
\bea
H^2 \, &=& \, - \frac{{\bar k}}{a^2} - \frac{\Lambda_E}{3} - \frac{2 {\bar k}^2 (\zeta + 3 \eta)}{\alpha a^4}
+ \frac{\rho}{6 \alpha} \, , \label{Heq} \\
\bigl[ {\dot H} + \frac{3}{2} H^2 \bigr] \, &=& \, - \frac{{\bar k}}{2 a^2} - \frac{\Lambda_E}{2} + 
\frac{{\bar k}^2 (\zeta + 3 \eta)}{4 \alpha a^4} + \frac{p}{4 \alpha} \, , \label{acceq}
\eea
and
\be
{\ddot \varphi} + 3 H {\dot \varphi} + V^{\prime} \, = \, 0 \, , \label{mattereq}
\ee
where $H = {\dot a}/a$, $p$ and $\rho$ are the pressure and energy
density of the scalar matter field, respectively, a prime denotes the derivative
with respect to $\varphi$, the dimensionless constant $\eta$
\be
\eta \, = \, \frac{\kappa^2 \mu^2 (1 - 4 \lambda)}{32 (1 - 3 \lambda)}
\ee
is the coefficient of the $R^2$ term in the gravitational action, and the
dimensionless constant $\zeta$ is given by
\be
\zeta \, = \, - \frac{\kappa^2 \mu^2}{8} \, .
\ee

The key new term in the cosmological equations of motion is the
second to last term on the right-hand sides of (\ref{Heq}) and
(\ref{acceq}). This term corresponds to ``dark radiation" with
a negative energy density. This term is present only if the spatial
curvature of the metric is non-vanishing. If the energy density of
regular matter increases less fast than $a^{-4}$ as the scale
factor decreases, the dynamics will lead to a cosmological
bounce provided that \footnote{The contributions scaling like
curvature or a cosmological constant can be neglected.}
\be \label{cond}
(\frac{\rho}{12} - p) \, > \, 0 \, .
\ee

To obtain a bounce in Ho\v{r}ava-Lifshitz cosmology we will
assume that matter in the pre-bounce epoch is described
by a scalar field $\varphi$ with a potential
\be
V(\varphi) \, = \, \frac{1}{2} m^2 \varphi^2 \, .
\ee
As studied in detail in quintom bounce 
\cite{quintom,quintombounce,LWbounce}
scenarios, we take the scalar field to be oscillating during the
contracting phase, with an amplitude ${\cal A}(t)$. 

As the universe contracts, the amplitude ${\cal A}(t) \sim a(t)^{-3/2}$ will
increase. Once the amplitude reaches the value 
\be
{\cal A}_{crit} \, = \, (12 \pi)^{-1/2} m_{pl} \, ,
\ee
where $m_{pl}$ is the Planck mass, then the field oscillations
will stop and $\varphi$ will enter a ``slow climb" phase, the
time reversal of the inflationary slow-roll phase. During this
phase, the matter energy density is approximately constant
but the scale factor is rapidly decreasing. Hence, the
dark radiation term in the Hubble equation rapidly catches
up with the matter energy term. Since in the slow
climb phase the pressure of matter is negative, the
condition (\ref{cond}) is satisfied. Thus, we obtain a cosmological
bounce.

Note that the ``slow climb" phase is unstable with respect
to the presence of the second mode in the scalar field
equation of motion, a mode which is exponentially
decaying in an inflationary slow-roll phase and thus ensures
that the slow-roll trajectory in large-field inflation is a local
attractor \cite{Kung}. In the ``slow climb" phase, the second
mode is increasing. Thus, the ``slow climb" trajectory is
a repeller \footnote{We thank Misao Sasaki and Takahiro
Tanaka for discussions on this point.}. However, coming
out of the oscillatory phase, the initial amplitude of the
unstable mode is sufficiently small such that the instability
does not have time to develop before the bounce takes place.

\section{Fluctuations in the Matter Bounce Scenario}

In the following we assume that the contracting phase before the
bounce was first dominated by cold matter, matter with an equation
of state $w = 0$, where $w$ is the ratio of pressure $p$ divided by
energy density $\rho$. We will now review how an initial vacuum
spectrum of cosmological perturbations on sub-Hubble scales
in the contracting phase develops into a scale-invariant spectrum
for wavelengths which exit the Hubble radius in the matter-dominated
phase.

As has proven to be convenient in inflationary cosmology, we 
track the cosmological fluctuations
in terms of the variable ${\cal R}$, the curvature fluctuations in
co-moving coordinates \cite{Bardeen,BST,BK,Lyth} . This variable 
is conserved at phase transitions and is constant on super-Hubble 
scales in an expanding universe. 

If we work in longitudinal gauge in which the
metric in the absence of anisotropic stress takes the form
\be
ds^2 \, = \, a^2(\eta) \bigl[ (1 + 2 \Phi) d\eta^2 - (1 - 2 \Phi) d{\bf{x}}^2 \bigr] \, ,
\ee
where $\eta$ is conformal time, ${\bf{x}}$ are co-moving spatial
coordinates) and  $\Phi({\bf{x}}, \eta)$ describes the metric fluctuations,
then ${\cal R}$ is given by (modulo terms which are suppressed
on super-Hubble scales)
\be
{\cal R} \, = \, \frac{2}{3} \bigl( {\cal{H}} \Phi^{'} + \Phi \bigr) \frac{1}{1 + w} + \Phi \, ,
\ee
$\cal{H}$ denoting the Hubble expansion rate in conformal time and a prime
indicating the derivative with respect of $\eta$.

The variable ${\cal R}$ is closely related to the variable $v$
(see \cite{MFB} for an in depth review of the theory of cosmological
fluctuations and \cite{RHBrev1} for an introductory overview) 
in terms of which the action for cosmological
fluctuations has canonical kinetic term: 
\be \label{zeta}
{\cal R} \, = \, \frac{v}{z} \,
\ee
where $z$ is a function of the background which is proportional to the scale
factor $a$ as long as the equation of state of matter in constant.

The equation of motion for the Fourier mode $v_k$ of $v$ is
\be
v_k^{''} + \bigl( k^2 - \frac{z^{''}}{z} \bigr) v_k \, = \, 0 \, .
\ee
This shows that on length scales larger than the Hubble radius,
where the $k^2$ term is negligible, $v$ does
not oscillate, its time evolution being determined by the
the gravitational background, whereas on sub-Hubble scales $v_k$ is oscillating
with approximately constant amplitude.

On super-Hubble scales, the equation of motion for $v_k$ in a universe
which is undergoing matter-dominated contraction is 
\be
v^{''}_k \, = \, 2 \eta^{-2} v_k \, ,
\ee
which has the general solution
\be
v_k(\eta) \, = \, c_1 \eta^{-1} + c_2 \eta^{2} \, ,
\ee
where $c_1$ and $c_2$ are constants. Since for a matter dominated
phase 
\be
a(\eta) \, \sim \, \eta^2
\ee
it follows that the $c_2$ mode is the mode for which
${\cal R}$ is constant on super-Hubble
scales, whereas for the $c_1$ mode ${\cal R}$ scales
as $\eta^{-3}$, which is decaying in an expanding
universe but growing in a contracting phase. It is
this growth which is responsible for turning an initial
vacuum spectrum of fluctuations into a scale-invariant one.

To see this, let us compute the power spectrum of
${\cal R}$ on super-Hubble scales late in the contracting
phase:
\bea
P_{{\cal R}}(k, \eta) \, &\sim& \, k^3 |v_k(\eta)|^2 a^{-2}(\eta) \\
&\sim& \, k^3 |v_k(\eta_H(k))|^2 \bigl( \frac{\eta_H(k)}{\eta} \bigr)^2 \, 
\sim \, k^{3 - 1 - 2} \nonumber \\
&\sim& \, {\rm const}  \, . \nonumber 
\eea
In this first step, we have used the definition of the power
spectrum, replaced ${\cal R}$ by $v$ via (\ref{zeta}) and used
the scaling $z(\eta) \sim a(\eta)$. In the second step, we
made use of the growth of the $c_1$ mode, the dominant
mode, to relate the value of $v$ at late times to its value
at the time $\eta_H(k)$ when the mode $k$ crosses the Hubble radius.
Finally, in the last step we insert the vacuum spectrum for $v$ on
sub-Hubble scales and the 
Hubble radius crossing condition $\eta_H(k) \sim k^{-1}$.

\section{Evolving Fluctuations through the Bounce}

If the bounce phase is short compared to the wavelength of the
fluctuations which are being followed, then the spectrum of ${\cal R}$
is unchanged through the bounce. This result can be obtained by
explicitly evolving fluctuations through a non-singular bounce using
the equations of motion for fluctuations which follow from Einstein's
theory (see e.g. \cite{Omidbounce,Tirthobounce,quintombounce,LWbounce}).
This result also agrees with what is obtained by replacing the bounce phase
by a matching surface and making use of the Hwang-Vishniac \cite{HV}
(Durrer-Mukhanov \cite{DM}) matching conditions.

If the energy density at the bounce is of the order of $(10^{16} {\rm GeV})^4$,
the wavelength of a mode which corresponds to the current Hubble
radius is about $1 {\rm mm}$, i.e. in the far IR. Since in the Ho\v{r}ava-Lifshitz
bounce, the bounce time is set by the UV scale, it is well justified to assume
that in the context of the use of the Einstein equations for the gravitational
fluctuations the spectrum of ${\cal R}$ does not change across the bounce.

In addition, again since the scales we are interested in are in the far IR,
it should be justified to use the IR limit of Ho\v{r}ava-Lifshitz cosmology to
propagate the fluctuations \footnote{This claim should be justified with
an explicit calculation in the same way that the corresponding claim
was justified \cite{Biswas2} in the bouncing scenario obtained by
using the special higher derivative gravitational action of \cite{Biswas1}
which is ghost-free about Minkowski space-time.}

Thus, we argue that the scale-invariance of the spectrum of cosmological
perturbations will be scale-invariant after the bounce. Since ${\cal R}$ is
constant on super-Hubble scales in the post-bounce expanding phase,
it then immediately follows that the spectrum of fluctuations at late times
will be scale-invariant.

A bouncing cosmology in the context of Ho\v{r}ava-Lifshitz gravity can
thus provide an alternative to inflation for providing a scale-invariant
spectrum of cosmological perturbations, provided that we begin in
the contracting phase with quantum vacuum fluctuations and provided
that the relevant scales exit the Hubble radius in a period of cold matter
domination (the case of initial thermal fluctuations in analyzed in
\cite{thermalbounce}). 

Since the curvature perturbation ${\cal R}$ grows on super-Hubble
scales, a matter bounce leads to a larger amplitude of non-Gaussianities
than slow-roll single-field inflation. Since it is a
different mode of ${\cal R}$ which dominates, the shape of the non-Gaussianities
is also different from what is obtained in slow-roll single-field inflation models.
The specific predictions for the amplitude and shape of the three-point function
(the ``bispectrum") were worked out in \cite{matterbounceng}. In particular,
the predicted amplitude of the bispectrum is very close to the level which
could be detected using the Planck satellite experiment.

\section{Conclusions and Discussion}

In this paper we have shown how to obtain a ``matter bounce" if Ho\v{r}ava-Lifshitz
gravity. Such a bounce is obtained because of a ``dark radiation" term which
appears in the equations of motion for cosmological solutions, a term which stems
from the terms in the quantum gravity theory which appear in the UV and help
render the theory renormalizable. Note, however, that the presence of the dark
radiation term requires non-vanishing spatial curvature. 

To obtain a cosmological bounce it is important that no source of matter is
present which redshifts equally fast or faster than that of the dark radiation
term. Only in this case the energy density of dark radiation can grown with respect
to the regular matter energy, a condition which is required to obtain a bounce.
This condition appears to be rather restrictive since it even rules out regular
radiation before the bounce.

We have presented a model in which a bounce can be obtained. In this model,
matter is modeled by a scalar field with a standard mass term. The scalar field
oscillates at early times in the contracting phase, leading to a matter equation of
state which is that of cold matter. Once the amplitude of scalar field oscillations
reaches a critical value, the field enters a deflationary slow-climb phase during
which its energy density is approximately constant and its pressure is negative.
Hence, a bounce occurs. 

In the matter bounce model thus constructed, initial quantum vacuum fluctuations
which exit the Hubble radius in the contracting matter-dominated phase
acquire a scale-invariant spectrum of curvature fluctuations, as already
envisioned in \cite{Wands,FB2,Wands2} and recently studied in detail in
\cite{LWbounce}. Thus, one of the main messages of this Note is that it is
not necessary to force a period of inflationary expansion into Ho\v{r}ava-Lifshitz
cosmology. The alternative matter bounce scenario predicts an amplitude
of the normalized bi-spectrum is the order of $1$, and a specific shape of this
three-point function, as studied in detail in \cite{matterbounceng}. These
specific predictions are potentially within the reach of upcoming CMB missions
such as PLANCK.

To obtain a successful late-time cosmology, the model presented here must
be supplemented with a mechanism to transfer the energy at late times
to Standard Model matter and radiation. If we include such matter in the
basic Lagrangian, then an initial condition problem arises: in order for
a bounce to occur, the initial energy density of radiation must be so low
that the radiation never comes to dominate during the contracting phase.
All of these issues deserve further study.

While this Note was being prepared for submission, a very interesting
paper by Mukohyama \cite{Mukh} appeared showing that in the UV region,
fluctuations of a scalar field in Ho\v{r}ava-Lifshitz gravity acquire a 
scale-invariant spectrum, a spectrum which can be later transformed
to curvature fluctuations via a transfer from entropy to adiabatic modes
such as the curvaton mechanism. Thus, one can obtain
a scale-invariant spectrum of cosmological fluctuations. An
advantage of this mechanism is that it also operates in a background
cosmology without a bounce (e.g. in the case of zero spatial curvature). However,
in the case of a matter bounce background, it is unclear whether the scaling
of the correlation functions used in \cite{Mukh} would extend to the large IR
scales required to match with observations, the scales on which our
mechanism works nicely.

\begin{acknowledgments}

The author wishes to thank G. Calcagni, P. Ho\v{r}ava and  S. Mukohyama for
comments on the draft of this Note.
This work is supported by an NSERC Discovery Grant and by the Canada Research
Chairs Program. The author wishes to acknowledge the hospitality of the Institute
of High Energy Physics in Beijing, where a lot of the initial work on the matter
bounce scenario was developed, and of the CERN Theory Division, where the
ideas presented here were worked out. 

\end{acknowledgments}

\end{document}